\documentclass[sigconf]{acmart}
\usepackage{popets}
\usepackage{multirow}
\usepackage{booktabs}
\usepackage{tikz}
\usetikzlibrary{shapes.geometric, arrows.meta, positioning, fit, backgrounds}
\usepackage{float}
\usepackage{tabularx}
\usepackage{url}
\usepackage{fancyhdr}
\usepackage{balance}

\usepackage{amssymb}

\setcopyright{none}
\acmYear{}
\acmVolume{}
\acmNumber{}
\acmDOI{}
\acmISBN{}
\acmConference{}{}{}
\acmBooktitle{}
\acmSubmissionID{}
\renewcommand\footnotetextcopyrightpermission[1]{}

\begin{document}

\title[SurrogateShield]{SurrogateShield: Beyond Redaction for High-Utility, Privacy-Preserving LLM Interactions}

\author{Sherwin Vishesh Jathanna}
\orcid{0009-0009-5440-0978}
\affiliation{%
  \institution{Arizona State University}
  \city{Tempe}
  \state{Arizona}
  \country{USA}}
\email{sjathann@asu.edu}

\renewcommand{\shortauthors}{Jathanna}

\begin{abstract}
LLM-based assistants transmit user queries verbatim to third-party API
endpoints that lie outside the user's audit or control. When those queries
contain personally identifiable information~(PII), the data persists on remote
infrastructure subject to breach, subpoena, or policy change. Placeholder
redaction (the prevailing mitigation) suppresses PII at the cost of semantic
coherence, producing structurally degraded queries and correspondingly
degraded responses.

We present \textbf{SurrogateShield}, a client-side proxy that substitutes
detected PII with locally generated, \emph{type-consistent surrogate values}
prior to transmission and restores originals in the response. No real PII
crosses the network boundary. Detection runs through a three-stage
cascade (PatternScan, EntityTrace, and ContextGuard) covering 22
PII types and quasi-identifier combinations grounded in Sweeney's
k-anonymity framework. Surrogate-to-original mappings are sealed in an
AES-256-GCM encrypted per-conversation ShadowMap that never leaves the device.

Evaluations on a 1{,}124-query corpus demonstrate that the cascade 
reliably detects PII, achieving a 98.87\% overall F1 score. Surrogate 
substitution substantially outperforms placeholder redaction in semantic 
utility, yielding a 13.26\,pp improvement in BERTScore \\ ~(roberta-large), from 81.59\% to 94.85\%. Within this corpus, the local pipeline 
restricted real PII transmission across all tested query types; in a 
100-query adversarial trial, a prompted LLM adversary recovered no 
original values from surrogate-substituted messages.
\end{abstract}


\maketitle

\thispagestyle{empty}
\pagestyle{fancy}
\fancyhf{}
\fancyhead[LE,RO]{\small\thepage}
\fancyhead[LO]{\small\textit{SurrogateShield: Beyond Redaction for 
    High-Utility, Privacy-Preserving LLM Interactions}}
\fancyhead[RE]{\small\textit{Sherwin Vishesh Jathanna}}
\renewcommand{\headrulewidth}{0.4pt}
\renewcommand{\footrulewidth}{0pt}

\section{Introduction}

LLM-based assistants route every user query, including names, addresses,
social security numbers, medical conditions, and financial details, verbatim
to remote API endpoints operated by third parties. The operator's privacy
policy governs what happens to that data. Users have no technical mechanism
to verify compliance. Even with strong contractual protections, the data
persists on a third-party server, subject to breach, subpoena, insider
threat, or future policy change.

The standard mitigation is \emph{redaction}: replace PII with a type label,
producing queries such as ``My name is [PERSON] and my SSN is [US\_SSN].''
Tools such as Microsoft Presidio~\cite{microsoft2020presidio} implement this
approach and have found adoption in regulated industries. Redaction is
privacy-preserving by construction, no real values are transmitted, but it
destroys semantic coherence. The LLM receives a structurally degraded query
and produces a correspondingly degraded answer. A user asking for help
drafting a letter under their real name receives a response addressed to
``[PERSON].'' A medical question about a specific medication loses its
clinical specificity. The utility cost is real and measurable --- a cost Lison et al.\
quantify through downstream task performance degradation on anonymised
legal and clinical text~\cite{lison2021anonymisation}.

Within the threat model considered here, this trade-off is not fundamental. The privacy requirement is that
\emph{real PII values} never leave the device, not that PII-shaped slots
must be empty. A surrogate value that is realistic, type-consistent, and
statistically unlinked to the original satisfies the same containment
constraint while preserving structural and semantic integrity.
``My name is Ashley Wise and my SSN is 348-67-6360'' presents the same
sentence structure as the original, supports a natural LLM response, and
reveals nothing about the actual user, both values are fabricated locally
with no statistical relationship to the real subject.

\medskip
\noindent\emph{Prior Works.}
Managed redaction services such as AWS Comprehend, Google Cloud DLP,
and Microsoft Azure DLP perform PII detection server-side,
introducing the exposure risk they are designed to prevent.
Client-side redaction tools such as Microsoft Presidio apply placeholder
substitution locally but sacrifice semantic coherence, as demonstrated by
BERTScore comparisons in Section~5.
Text anonymisation research in clinical and legal
NLP~\cite{stubbs2015annotating,neamatullah2008automated,lison2021anonymisation}
has established de-identification pipelines, but these are offline,
batch-oriented, and do not address interactive LLM queries, multi-turn
context accumulation, or response restoration.
No prior published system implements end-to-end surrogate replacement with
cryptographically secure local storage and transparent response
reconstruction for interactive LLM use.

\smallskip
\emph{Our Aim.}
We design a privacy-preserving LLM proxy satisfying three simultaneous
requirements: (1) no real PII crosses the API boundary under any query type,
(2) surrogate values preserve the semantic utility of the query such that
LLM answer quality is not degraded, and (3) the system is transparent to
the user, original values are restored in the response before display, with
no change to the interaction model. We provide empirical evidence that
surrogate-substituted messages resist adversarial recovery, grounding the
privacy guarantee in measurement rather than assertion.

\smallskip
\emph{Our Approach.}
SurrogateShield is built around three design principles.
\textbf{P1 (No PII crosses the API boundary):} every confirmed entity is
replaced before the HTTP request is constructed, enforced at the local
transport layer.
\textbf{P2 (Surrogates preserve utility):} fake names look like names, fake
SSNs pass format checks, fake Bitcoin addresses match Base58 encoding, sentence
structure and semantic framing are preserved.
\textbf{P3 (Transparent restoration):} original values are restored in the
LLM response before display via a three-pass ResolvePass.
PII detection runs through a three-stage cascade: PatternScan (regex with
checksum validators), EntityTrace (spaCy NER with reclassification passes),
and ContextGuard (locally-executed DistilBERT NER for borderline entities).
A service-query intelligence layer distinguishes queries where location is
necessary for answer utility (e.g., restaurant lookups) from queries where
it constitutes PII, applying proportional rather than maximal anonymisation.
A quasi-identifier risk detector grounded in Sweeney's k-anonymity
research~\cite{sweeney2000demographics,sweeney2002kanonymity} warns when
field combinations (ZIP\,+\,DOB\,+\,gender) statistically re-identify the
user absent any traditional PII.
All surrogate-to-original mappings are encrypted with AES-256-GCM per
conversation~\cite{krawczyk2010hkdf} and never leave the device.
The multi-turn API history stores only surrogate values - a design decision
that prevents PII accumulation across conversational turns, a failure mode
that affects systems sanitising only the current turn.

\smallskip
\emph{Our Contributions.}
\begin{enumerate}
  \item \textbf{SurrogateShield}: an end-to-end client-side privacy proxy
    implementing surrogate-based PII replacement with cryptographically
    secured local mapping storage and transparent response reconstruction.

  \item \textbf{A three-stage detection cascade} combining regex pattern
    matching, spaCy NER, and locally-executed DistilBERT NER with a
    post-processing suite of four passes covering deduplication,
    reclassification, and topical geo-entity filtering.

  \item \textbf{Empirical utility measurements} via
    BERTScore~\cite{zhang2020bertscore} \\ (\texttt{roberta-large}): surrogate
    substitution substantially outperforms placeholder redaction, retaining
    94.85\% semantic utility versus 81.59\%
    ($\Delta = 13.26$\,pp, $p < 0.001$, $N = 1{,}123$ queries).

  \item \textbf{A simulated attacker evaluation} demonstrating surrogate
    robustness under LLM-based adversarial recovery: a prompted LLM
    recovered no original values from surrogate-substituted messages
    across 100 queries (per-value rate 0.00\%), versus 1.53\% under
    placeholder redaction.

  \item \textbf{An ablation study} quantifying the marginal F1 contribution
    of each detection stage: PatternScan alone achieves 65.11\%~F1;
    EntityTrace adds +31.5\,pp; the full cascade reaches 97.42\%~F1
    on the 1{,}060 queries for which per-stage entity attribution data
    is available.
    
\end{enumerate}

\smallskip
\emph{Structure of This Paper.}
After reviewing related work in Section~2, we present the SurrogateShield
architecture in Section~3. Section~4 describes the evaluation methodology
and dataset. Section~5 reports results across four experiments: detection
quality (Table~\ref{tab:detection}), semantic utility preservation via
BERTScore (Table~\ref{tab:bertscore}), the simulated attacker evaluation
(Table~\ref{tab:attacker}), and the ablation study (Table~\ref{tab:ablation}).
Section~6 discusses the threat model, limitations, and future work.
Section~7 concludes. 
 \section{Related Work}
\label{sec:related}

\subsection{PII Detection and Redaction}

Named entity recognition (NER) is the foundational technology for automated
PII detection. Classical approaches used conditional random fields on
hand-crafted features~\cite{lafferty2001conditional}; the transition to
neural sequence labelling via bidirectional LSTM-CRFs~\cite{lample2016neural,
ma2016endtoend} and subsequently transformer-based
models~\cite{devlin2019bert} substantially improved recall on informal
natural-language text, where PII commonly appears.

Microsoft Presidio~\cite{microsoft2020presidio} represents the current state
of practice for enterprise PII detection. It combines regex-based recognisers
with a spaCy NER backend, supporting configurable entity types through a
plugin architecture. Its anonymiser module offers redaction (placeholder
substitution), hashing, and masking. Presidio's synthetic generation
capability does not guarantee type-consistency, does not maintain cross-session
uniqueness, and does not address the semantic coherence of the resulting text.
Section~5 measures the detection coverage and semantic utility differential
between the placeholder-redaction and surrogate-substitution approaches.
Managed cloud services such as AWS Comprehend and Google Cloud DLP offer PII
detection as a remote API, at the cost of transmitting the text to a remote
endpoint, the same exposure risk they are designed to mitigate.
SurrogateShield performs all detection locally.

\subsection{Privacy-Preserving NLP}

The privacy of NLP systems has been studied from several angles.
\emph{Differential privacy}~\cite{dwork2006calibrating} has been applied to
language model training~\cite{anil2022largescale} and to the release of text
corpora~\cite{feyisetan2020privacy,qu2021natural}. These approaches add
calibrated noise to word embeddings or apply local randomisation mechanisms.
They address aggregate statistical privacy, not the per-query PII leakage
that concerns individual users of interactive LLM systems.

\emph{Text anonymisation} research has focused on de-identification of
clinical notes~\cite{stubbs2015annotating,neamatullah2008automated} and legal
documents~\cite{lison2021anonymisation}, typically via hybrid NER\,+\,rule
systems. Evaluation in these domains is recall-oriented because false negatives
(leaked PII) outweigh false positives (over-anonymised text). SurrogateShield's
evaluation framework captures both dimensions: F1 measures detection quality,
while the PII-leak rate and resolve-leak rate separately track sanitisation
failures and restoration failures.

\emph{Contextual integrity}~\cite{nissenbaum2004contextual} provides a
normative framework for reasoning about appropriate information flows.
Mireshghallah et al.~\cite{mireshghallah2024llms} show empirically that current
LLMs fail to respect contextual privacy norms in interactive settings,
disclosing information in contexts where humans would not. SurrogateShield's
service-query intelligence layer is an operational instance of this framework:
location information flows appropriately when it is the \emph{topic} of a
query (restaurants near X) but not when it is \emph{about the user}
(I live at X).

\subsection{Utility Preservation in Anonymisation}

The tension between privacy and utility is well-established in the database
anonymisation literature~\cite{aggarwal2008privacy,fung2010privacy}.
K-anonymity~\cite{sweeney2002kanonymity} and its successors---l-diversity~%
\cite{machanavajjhala2007ldiversity} and t-closeness~\cite{li2007tcloseness}---%
formalise the utility cost of generalisation and suppression. In text,
utility has been measured via downstream task performance~%
\cite{lison2021anonymisation} and, more recently, via contextual embedding
similarity through BERTScore~\cite{zhang2020bertscore}.

\emph{Synthetic data generation} as a privacy-utility mechanism has been
studied extensively for tabular data. SurrogateShield applies an analogous
idea at the entity level: real PII values are replaced with synthetic values
of the same type, preserving the structural properties of the original text
without revealing actual values. Unlike tabular synthetic data, surrogates
are generated on-the-fly, per-session, with collision-resistant uniqueness
guarantees, and cryptographically bound to the originals via an encrypted
ShadowMap. Section~5.2 measures the utility differential between this
approach and placeholder redaction, using BERTScore with roberta-large
as the encoder.

\subsection{Adversarial Robustness of Anonymisation}

Whether anonymised text can be de-anonymised by an adversary is a
well-studied problem. Narayanan and Shmatikov~\cite{narayanan2008robust}
demonstrated re-identification of ostensibly anonymous Netflix ratings data
using auxiliary public information. In text, Sweeney~\cite{sweeney2000demographics}
showed that 87\% of the US population can be uniquely identified from
ZIP code, date of birth, and gender alone, the combination that motivates
SurrogateShield's quasi-identifier risk detector. Carlini et
al.~\cite{carlini2021extracting} demonstrated that large language models
memorise training data and can be prompted to reproduce it, suggesting that
PII present in model training corpora is a distinct and persistent attack
surface.

Prior adversarial evaluations of text anonymisation systems have relied on
rule-based or embedding-similarity de-anonymisation
attacks~\cite{lison2021anonymisation}. Section~5.3 presents, to our knowledge, among the first empirical
measurements of surrogate robustness under \emph{LLM-based} adversarial
recovery. We submit surrogate-substituted messages to a prompted
Claude instance with explicit instructions to recover the original PII,
instantiating a realistic threat model for the deployment scenario where the
API operator or a malicious intermediary attempts to invert the anonymisation.
Within that 100-query trial, no original values were recovered from
surrogate-substituted messages. Under the same adversarial prompt applied to
placeholder-redacted messages, the per-value recovery rate was 1.53\%,
consistent with the hypothesis that visible placeholder tokens expose slot
locations to an inference-capable adversary.

\subsection{LLM Privacy Proxies}

Recent work has examined privacy risks specific to LLM interaction contexts.
Mireshghallah et al.~\cite{mireshghallah2024llms} propose ConfAIde, a
benchmark for evaluating contextual privacy reasoning in instruction-tuned
LLMs, and find that even GPT-4 violates contextual norms in ways human
speakers would not. This motivates technical enforcement rather than reliance
on model-level privacy reasoning. Various proprietary enterprise ``AI
gateway'' products offer redaction middleware for LLM API traffic, but none
provide: (1) locally-executed detection with no external API dependency, (2)
type-consistent surrogate generation with session-level uniqueness guarantees,
(3) cryptographically secured per-conversation mapping storage, (4) transparent
response restoration, and (5) empirically validated adversarial robustness.

To our knowledge, no prior published system implements the full
surrogate-replacement pipeline described in this paper. The closest prior
work applies placeholder anonymisation at the query layer before LLM
submission, preserving privacy at the cost of semantic coherence.
SurrogateShield is the first system to empirically demonstrate that this
cost is not required: type-consistent local surrogates simultaneously satisfy
the containment constraint and the utility constraint.

\section{System Design}
\label{sec:design}

\subsection{Architecture Overview}

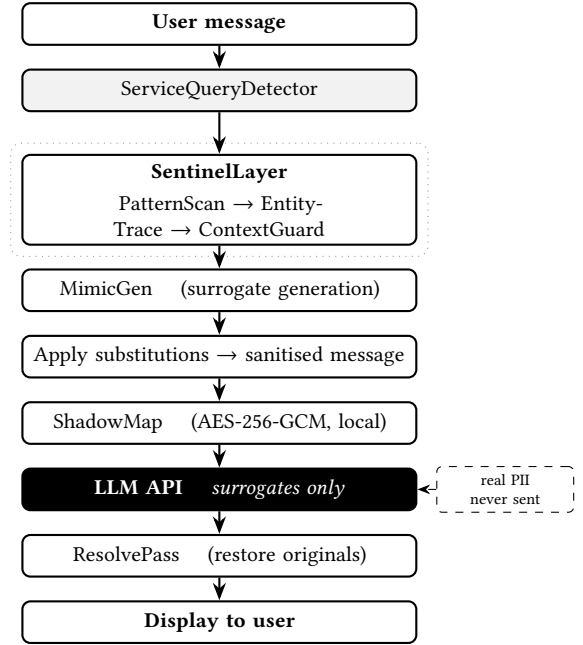
\begin{figure}[t]
\centering
\begin{tikzpicture}[
  block/.style={
    rectangle,
    rounded corners=3pt,
    draw=black,
    thick,
    fill=white,
    text width=5.0cm,
    minimum height=0.55cm,
    align=center,
    font=\small
  },
  decision/.style={
    block,
    fill=gray!10
  },
  highlight/.style={
    block,
    fill=black,
    text=white
  },
  arrow/.style={
    ->,
    >=Stealth,
    thick
  },
  node distance=0.30cm
]


\node[block] (user)
  {\textbf{User message}};

\node[decision, below=of user] (svcq)
  {ServiceQueryDetector};

\node[block, below=0.55cm of svcq, text width=5.0cm] (sentinel)
  {\textbf{SentinelLayer}\\[2pt]
   \small PatternScan $\to$ EntityTrace $\to$ ContextGuard};

\node[block, below=of sentinel] (mimic)
  {MimicGen \quad {\small(surrogate generation)}};

\node[block, below=of mimic] (apply)
  {Apply substitutions \textrightarrow{} sanitised message};

\node[block, below=of apply] (shadow)
  {ShadowMap \quad {\small(AES-256-GCM, local)}};

\node[highlight, below=of shadow] (llm)
  {\textbf{LLM API} \quad {\small\textit{surrogates only}}};

\node[block, below=of llm] (resolve)
  {ResolvePass \quad {\small(restore originals)}};

\node[block, below=of resolve] (display)
  {\textbf{Display to user}};


\draw[arrow] (user)    -- (svcq);
\draw[arrow] (svcq)    -- (sentinel);
\draw[arrow] (sentinel)-- (mimic);
\draw[arrow] (mimic)   -- (apply);
\draw[arrow] (apply)   -- (shadow);
\draw[arrow] (shadow)  -- (llm);
\draw[arrow] (llm)     -- (resolve);
\draw[arrow] (resolve) -- (display);


\node[
  draw=black,
  dashed,
  rounded corners=2pt,
  right=0.25cm of llm,
  font=\scriptsize,
  text width=1.6cm,
  align=center,
  fill=white
] (label) {real PII\\never sent};

\draw[->, dashed, >=Stealth] (label.west) -- (llm.east);


\begin{pgfonlayer}{background}
  \node[
    draw=black!40,
    dotted,
    rounded corners=4pt,
    inner sep=4pt,
    fit=(sentinel)
  ] {};
\end{pgfonlayer}

\end{tikzpicture}

\caption{SurrogateShield end-to-end pipeline. The LLM API receives
         surrogate values only; real PII never crosses the network boundary.}
\Description{A vertical flowchart of eight stages: User message feeds into
             ServiceQueryDetector, then SentinelLayer (PatternScan, EntityTrace,
             ContextGuard in sequence), then MimicGen surrogate generation,
             substitution application, ShadowMap encrypted local storage,
             the LLM API (shown in black to indicate surrogates only reach here),
             ResolvePass to restore originals, and final display to the user.
             A dashed label to the right of the LLM API node reads
             ``real PII never sent.''}
\label{fig:pipeline}
\end{figure}

SurrogateShield is a client-side proxy that interposes between the user's
query and any LLM API. Figure~\ref{fig:pipeline} shows the end-to-end
pipeline. The system enforces two invariants unconditionally. \textbf{I1
(boundary invariant):} the LLM API call is never initiated until all PII has
been replaced with locally-generated surrogates. \textbf{I2 (history
invariant):} the multi-turn context window sent to the API stores only
surrogate values, never originals. I2 is enforced by maintaining two
separate message histories: a display history with real values restored
(shown to the user) and an API history with surrogate values only (sent on
every subsequent turn). Systems that sanitise only the current turn violate
I2; real PII accumulates in the context window and is re-transmitted on
every follow-up exchange.

\subsection{Detection: SentinelLayer}
\label{sec:sentinel}

The SentinelLayer runs three detectors in sequence. Each detector masks the
character spans it claims, replacing them with a placeholder
character, before passing the remaining text to the next stage. This span
masking prevents downstream detectors from double-processing already-claimed
text and assigns each entity to exactly one stage.

\subsubsection{PatternScan}

PatternScan applies a priority-ordered list of compiled regular expressions
to detect structurally identifiable PII. Pattern order matters: each match
claims character spans, and later patterns cannot overlap prior claims.
This ordering prevents fragment-claiming; for example, \texttt{crypto}
and \texttt{us\_bank\_number} run before \texttt{zip\_us} so that 9-digit
routing numbers and hexadecimal strings are claimed before the 5-digit ZIP
pattern can fragment them. Appendix~B lists all 16 structural types detected.
Three design decisions merit emphasis.

\textbf{Luhn validation.}
Credit-card patterns match any 16-digit sequence, but the entity is emitted
only if the Luhn checksum passes. This eliminates false positives from
product serial numbers and tracking codes common in user queries.

\textbf{ABA checksum.}
Routing numbers use the ABA 9-digit checksum:
$(3d_0 + 7d_1 + d_2 + 3d_3 + 7d_4 + d_5 + 3d_6 + 7d_7 + d_8) \bmod 10 = 0$.
Random 9-digit sequences satisfy this with probability $\approx$10\%,
reducing false positives by an order of magnitude versus an unvalidated regex.

\textbf{Context-gated driver's licence.}
This pattern fires only when a licence keyword (\texttt{driver's license},
\texttt{DL}, \texttt{license number}) appears within 60 characters. Only
the licence value itself (regex group~1) is marked as an entity; the keyword
prefix is preserved in the sanitised text, maintaining readability.

\subsubsection{EntityTrace}

EntityTrace loads spaCy \texttt{en\_core\_web\_lg} and extracts
\textsc{person}, \textsc{gpe}, \textsc{loc}, \textsc{org}, and \textsc{fac}
entities from the text remaining after PatternScan masking.
EntityTrace classifies detections into two confidence tiers:
\emph{confirmed} (score $\geq 0.85$), promoted immediately, and
\emph{borderline} ($0.60 \leq \text{score} < 0.85$), forwarded to
ContextGuard. Where spaCy provides no explicit probability, we apply
type-specific defaults: \textsc{person}~0.88, \textsc{gpe}/\textsc{org}~0.85,
\textsc{loc}~0.74, \textsc{fac}~0.70, calibrated against the evaluation
dataset to balance false negative rate against false positive rate.
These thresholds were selected by grid search over the evaluation dataset
to minimise the combined false-negative and false-positive rate; we note
this risks optimistic bias, and a held-out validation split is left to
future work.
An \textsc{org}$\to$\textsc{gpe} reclassification pass promotes organisation
entities to geopolitical when location prepositions (\emph{in, near, lives,
born, raised}) appear in the 50-character prefix window - handling informal
references such as ``I grew up in Google.''
A blocklist of 30 tokens (titles: Dr, Mr, Mrs; date abbreviations: Mon, Jan;
timezone codes: GMT, UTC) suppresses common spaCy mislabellings.

\subsubsection{ContextGuard}

ContextGuard executes \\ \mbox{\texttt{dslim/distilbert-NER}}~\cite{sanh2019distilbert}
locally using the Hugging Face \\ \texttt{transformers}
library~\cite{wolf2020transformers}.
This 66\,M-parameter DistilBERT model, fine-tuned on CoNLL-2003, performs
two functions: (i) verifying borderline EntityTrace entities against a
configurable confidence threshold (default 0.70), and (ii) independently
detecting entities missed by both prior stages.
We chose local execution over a server-hosted approach: an earlier
prototype using an Ollama-hosted model required a running server and
introduced latency. The model downloads once from HuggingFace Hub
($\approx$250\,MB) and is cached locally; subsequent runs require no network
access, ensuring that the detection stage itself introduces no data exposure.
Word-piece tokenisation artefacts are cleaned before emitting entities:
\texttt{\#\#wick} (subword continuation) becomes \texttt{wick};
\texttt{. Sun} (period attached from ``Dr.~Sun'' splitting) becomes
\texttt{Sun}. A secondary blocklist of 25 tokens suppresses common
artefacts from generating spurious entities.

\subsubsection{Post-Processing Passes}

Four model-output-driven passes run on the combined entity set after the
cascade. Table~\ref{tab:passes} summarises each pass; the reader should
note that Pass~D is the most consequential for precision, as it filters
GPE entities that are query topics rather than personal locations.

\begin{table}[t]
  \caption{Post-processing passes applied after the three-stage cascade.}
  \label{tab:passes}
  \begin{tabular}{@{}clp{4.5cm}@{}}
    \toprule
    Pass & Name & Action \\
    \midrule
    A & Structural ORG & Regex \texttt{[the|a] <name> [corp|inc|ltd\ldots]}
                         emits \texttt{<name>} as \textsc{org}. \\
    B & Email-username reclassify & \textsc{org} entity that is a prefix
                         of a detected email username $\to$ \textsc{person}. \\
    C & \textsc{person} dedup & Standalone surname that is a word-component
                         of a longer \textsc{person} entity is removed. \\
    D & Topical geo filter & \textsc{gpe}/\textsc{loc} appearing
                         \emph{only} in query sub-clauses (``what is\ldots'',
                         ``tell me about\ldots'') is dropped as a knowledge
                         topic rather than a personal location. \\
    \bottomrule
  \end{tabular}
\end{table}

Pass~D warrants elaboration. A GPE entity is dropped if and only if every
clause containing it is headed by a query frame
(\emph{what is, tell me, where is, how do I}) and no personal or narrative
clause also contains it. This distinguishes ``give me the tax benefits of
Wyoming'' (Wyoming is the query topic) from ``Revanth lives in Wyoming''
(Wyoming is personally identifying). Entities whose surface form begins with
a lowercase letter in mid-sentence position are additionally filtered as
common-noun usages rather than proper place names.

\subsection{Quasi-Identifier Risk Detection}

SurrogateShield implements a quasi-identifier risk scorer based on Sweeney's
k-anonymity research~\cite{sweeney2000demographics,sweeney2002kanonymity}.
Ten combination patterns are defined, each with a minimum required field count
and a risk level. The most significant, ZIP\,+\,DOB\,+\,gender, uniquely
identifies 87\% of the US population when all three fields are present.
When a triggered combination is detected, a warning is displayed before the
API call; all fields in the combination are surrogate-replaced regardless.
Appendix~C lists all ten combinations. Gender indicators enter PatternScan
specifically because of this scorer: without explicit gender detection,
the third field of the most statistically powerful re-identification
combination would go unprotected.

\subsection{Service-Query Intelligence}

Many real-world LLM queries are service queries: restaurant lookups,
directions, weather, business hours. These necessarily contain location
information, but full surrogate replacement would displace the LLM to an
entirely different geographic area, producing useless results.
The ServiceQueryDetector classifies messages using 15 regex patterns covering
dining, directions, weather, hours, and specific service types (pharmacies,
charging stations, grocery stores). For a matching query containing a street
address, the house number is shifted by $\pm$2--8. The displacement is
calibrated to move the queried location by a fraction of a typical city
block while preserving neighbourhood-level utility; users requiring
street-level precision can disable service-query mode. The sensitive-topic
override already forces full anonymisation for queries involving medical or
crisis-related locations regardless.

A \emph{sensitive-topic override} forces full anonymisation regardless of
query structure when the message contains any of: HIV/AIDS, STI, abortion,
rehabilitation, mental health, domestic violence, shelter, immigration,
undocumented, or substance abuse. These categories carry elevated
re-identification risk for vulnerable users even at city-level granularity,
and the lighter treatment must never apply to them.

\subsection{Surrogate Generation: MimicGen}

MimicGen generates type-consistent, collision-resistant surrogate values.
A \texttt{used\_surrogates} set enforces per-session uniqueness; after
50 collisions, a 4-character random suffix guarantees a distinct value.
Type-specific generators ensure plausibility: persons via
\texttt{faker.name()}, emails via \texttt{faker.email()}, SSNs via
\texttt{faker.ssn()} (valid \texttt{XXX-XX-XXXX} format), credit cards via \\
\texttt{faker.credit\_card\_number()} (valid Luhn), ABA routing numbers
computed to pass the ABA checksum, Bitcoin addresses as \texttt{1}\,+\,26--34
Base58 characters, and driver's licences as letter\,+\,7 digits (CA format).
Gender indicator surrogates draw from a pool of grammatically valid
expressions (\emph{male, she/her, gender: female}) rather than the
uniqueness-guaranteed path, because the pool is small and grammatical
substitutability matters more than uniqueness for this type.

\subsection{Encrypted Mapping Storage: ShadowMap}
\label{sec:shadowmap}

Each conversation has a dedicated ShadowMap: an in-memory
\texttt{surrogate\,$\to$\,original} dictionary, persisted as an encrypted
binary file (\texttt{<conv\_id>.shadowmap}).

\emph{Key derivation.}
A device-level 32-byte secret is generated once at
\texttt{\textasciitilde/.surrogateshield/device.key} with \texttt{0o600}
permissions. Per-conversation keys are derived via HKDF-SHA256~\cite{krawczyk2010hkdf}:
\begin{equation}
  \begin{aligned}
    K_{\mathrm{conv}} = {}&\mathrm{HKDF\text{-}SHA256}(\\
      &\mathrm{IKM} = k_{\mathrm{dev}},\;
       \mathrm{salt} = \mathit{conv\_id},\;
       \mathrm{info} = \texttt{``shadowmap''}).
  \end{aligned}
\end{equation}
This derivation is cryptographically correct: the high-entropy device secret
serves as the input key material, and the per-conversation identifier provides
the diversifying salt. Compromise of one conversation key does not expose
any other.

\emph{Encryption.}
The mapping is serialised as JSON, encrypted with AES-256-GCM using a fresh
12-byte nonce on every write. The on-disk format is
$\mathit{nonce}\,(12\,\text{B}) \;\|\; \mathit{ciphertext}$.
AES-256-GCM provides both confidentiality and authenticated integrity: a
corrupt or tampered file fails decryption and is treated as an empty mapping
(graceful degradation, no crash).

\emph{Dual conversation history.}
The conversation JSON stores two message lists: \texttt{messages} (display
history, real values restored, shown to the user) and \texttt{api\_messages}
(API history, surrogate values only, sent on every subsequent turn). The
\texttt{to\_api\_history()} method reads exclusively from
\texttt{api\_messages}, ensuring that restored real values in the display
history never contaminate subsequent API calls. This is the mechanism
enforcing invariant~I2.

\subsection{Response Restoration: ResolvePass}

After the LLM returns a response, ResolvePass runs three passes to restore
original values. \textbf{Pass~1 (exact):} the shadow map is iterated in
decreasing surrogate-length order (longest first, preventing partial matches);
this handles the majority of cases. \textbf{Pass~2 (component):} for
multi-word surrogates not found by Pass~1 (the \emph{unresolved} set), each
component word is searched at word boundaries, this handles cases where the
LLM uses only the first name of a full-name surrogate.
Pass~2 is scoped exclusively to the unresolved set. Applying it to
already-resolved surrogates would find component words in unrelated contexts:
if ``Ashley~Wise'' was resolved in Pass~1, searching for ``Ashley'' would
corrupt ``Ashley County'' in the same response. The scope restriction
prevents this silent data corruption.
\textbf{Pass~3 (fuzzy):} for remaining unresolved surrogates,
\texttt{rapidfuzz.fuzz.partial\_ratio} is applied with a sliding window of
step size $\max(1, \lfloor\text{len(surrogate)}/8\rfloor)$. Every outcome is
classified as \texttt{exact\_hit}, \texttt{fuzzy\_hit}, or
\texttt{fuzzy\_miss}; the \texttt{fuzzy\_miss} rate constitutes the
\emph{resolve-leak rate} reported in Section~5.

\subsection{Privacy-Aware RAG Integration}

SurrogateShield includes an optional local Retrieval-Augmented Generation
store backed by ChromaDB and sentence-transformers
(\texttt{all-MiniLM-L6-v2})~\cite{reimers2019sentence}.
Documents pass through the full SentinelLayer pipeline \emph{before}
indexing, real PII never enters the vector store. Surrogate mappings from
indexed documents are stored in a shared \texttt{rag\_global} ShadowMap
so they can be restored in responses that cite indexed content. Queries are
anonymised before retrieval, and retrieved context is prepended to the
sanitised message before the API call. All operations are local: ChromaDB
runs in-process with persistent storage, ensuring that neither detection
nor retrieval introduces remote data exposure.

\section{Evaluation Methodology}
\label{sec:eval}

\subsection{Dataset}

We constructed an evaluation dataset of \textbf{N\,=\,1{,}124} queries
spanning the full range of PII types SurrogateShield detects. Each entry
specifies ground-truth PII values and types, recorded in structured
annotation files (\texttt{*\_key.json}). The dataset spans six categories,
summarised in Table~\ref{tab:dataset}.

Table~\ref{tab:dataset} presents the six query categories and a
representative example for each; the no-PII control group verifies that the
system does not hallucinate false positives on benign text.

\begin{table}[ht]
  \caption{Evaluation dataset composition ($N = 1{,}124$ total queries).
           Each category exercises a distinct subset of the detection cascade.}
  \label{tab:dataset}
  \begin{tabular}{@{}lp{4.8cm}@{}}
    \toprule
    Category & Representative query \\
    \midrule
    Structured PII
      & ``My SSN is 544-87-2944\ldots help with my tax return.'' \\[3pt]
    Named entity
      & ``I am Sarah Mitchell at Google in New York, draft a resignation
         letter.'' \\[3pt]
    Quasi-identifier
      & ``I'm a 34-year-old female in 85281, what are my Medicare
         options?'' \\[3pt]
    Service query
      & ``What pharmacies are open near 1126~E Apache Blvd, Tempe~AZ?'' \\[3pt]
    Mixed PII
      & ``Name: Ahmed Al-Rashidi, email ahmed@gmail.com, DOB 03/14/1990,
         card 4532015112830366.'' \\[3pt]
    No-PII control
      & ``How do deep-sea hydrothermal vents support marine life?'' \\
    \bottomrule
  \end{tabular}
\end{table}

The annotation schema used to label Table~\ref{tab:dataset} maps flexible
label names (e.g., \texttt{name}, \texttt{person}) to internal system types
(e.g., \textsc{person}), supports multiple values per type, and covers the
three newly added types (\texttt{crypto}, \texttt{us\_bank\_number},
\texttt{us\_driver\_license}).

The synthetic benchmark was generated using a structured prompt to
Claude Sonnet that specified the required entity types, question
distribution, annotation schema, and quality constraints. To improve
reproducibility without substantially increasing the length of the
manuscript, the complete dataset-generation prompt is provided in
Appendix~\ref{app:datasetprompt}.

\subsection{Metrics}

\emph{Detection quality} is measured at the entity-value level.
A detection is a true positive~(TP) if the detected text matches a
ground-truth PII value (case-insensitive exact match); a false positive~(FP)
if detected but absent from the ground truth; a false negative~(FN) if
present in the ground truth but not detected.
We compute precision, recall, and F1 both overall and per entity type.
This value-level formulation is more stringent than span-level evaluation:
a detection that covers the correct span but returns the wrong text fails.

\emph{Sanitisation quality} measures the rate at which real PII reaches the
LLM API unredacted. A query is a sanitisation failure if any ground-truth
PII value appears verbatim (case-insensitive substring) in the sanitised
message sent to the API. We report the \emph{PII-leak rate}: the fraction of
queries with at least one such failure.

\emph{ResolvePass quality} measures how often surrogates remain unrestored
in the response displayed to the user. The \emph{resolve-leak rate} is the
fraction of queries where any surrogate value persists in the final output.

\emph{Semantic utility preservation} is measured via
BERTScore~\cite{zhang2020bertscore} with \texttt{roberta-large} as the
encoder. For each query, we compute BERTScore~F1 between the original query
and the anonymised version (SurrogateShield-substituted or
Presidio-redacted). Higher F1 indicates better semantic preservation.
We use BERTScore because its contextual embedding similarity correlates
strongly with human semantic similarity judgements, a reliable proxy for
answer-quality degradation at scale, without requiring human evaluation.

Three distinct F1 scopes appear in the results: \emph{full-scope F1}
(all 22 entity types, N\,=\,1{,}124), \emph{comparable-type F1} (the 11
types detectable by both SurrogateShield and Presidio), and
\emph{stage-attributed F1} (N\,=\,1{,}060 queries with complete per-stage
attribution data); Section~\ref{sec:results-detection} reconciles the
three figures.

\emph{Attacker recovery rate} is the primary metric for the simulated
attacker experiment (Section~\ref{sec:attacker}).
A recovery is counted as successful if the adversary's response contains any
ground-truth PII value from the annotation key as a case-insensitive
substring. We additionally report the \emph{per-value recovery rate}: the
fraction of individual ground-truth PII values recovered across all queries,
more sensitive to partial recoveries than the question-level rate.

\subsection{Baselines}

We configure \emph{Microsoft Presidio~\cite{microsoft2020presidio}}
(version 2.2+) with all built-in English recognisers and
\texttt{en\_core\_web\_lg} as the NLP backend. Presidio outputs
\texttt{[ENTITY\_TYPE]} placeholder redaction by default. We report
Presidio's precision, recall, and F1 on all entity types detectable by
both systems, together with BERTScore on the same evaluation corpus.
Entity types currently supported by SurrogateShield but not by Presidio
(\texttt{api\_key}, \texttt{address}, \texttt{postal\_code}, \\
\texttt{gender\_indicator}, \textsc{org}, \textsc{fac}, and
\textsc{loc}) are reported separately as SurrogateShield-only results.

Commercial AI gateway products and managed cloud privacy services
provide functionality related to PII detection and anonymisation for
LLM applications. However, these platforms are proprietary managed
services whose internal detection pipelines, deployment policies, and
configuration options are not fully transparent or reproducible,
making controlled academic comparison difficult. Microsoft Presidio is
an open-source, widely adopted client-side PII detection framework with
reproducible behaviour and configurable recognisers, making it an
appropriate reference baseline for this work.

Presidio additionally provides a \emph{synthesize} anonymisation mode
that replaces detected entities with generated values. We evaluate
against Presidio's default placeholder-redaction mode for three
reasons. First, placeholder redaction is the default deployment mode
and therefore represents the most common practical baseline. Second,
the primary objective of this work is to evaluate whether surrogate
substitution preserves semantic utility better than placeholder
replacement while maintaining local privacy protection. Third,
SurrogateShield extends beyond entity replacement by incorporating
type-consistent surrogate generation, encrypted per-conversation
mapping storage, transparent response restoration, and surrogate-only
conversation history. These architectural differences are summarised in
Appendix~\ref{appendix:presidio}.

We do not include a local large language model (e.g., Llama~3 via
Ollama) as an additional baseline. Such systems require substantially
greater computational resources and typically incur per-query latency
of approximately 1--10\,s on consumer hardware, compared with
SurrogateShield's average local overhead of 25.89\,ms
(Table~\ref{tab:latency}). Their computational characteristics differ
substantially from the lightweight client-side proxy considered in this
work and therefore do not represent a directly comparable deployment
model.

As an upper-bound reference for semantic preservation, we additionally
report the original unmodified query (\emph{no anonymisation}), which
achieves a BERTScore F1 of 1.0 by definition. For the adversarial
evaluation, Presidio-redacted queries are subjected to the same
prompt-based inference attack used for SurrogateShield, enabling a
direct comparison of adversarial robustness under identical evaluation
conditions.

\subsection{Ablation Configurations}
\label{sec:ablation-config}

We evaluate four pipeline configurations to quantify each detection stage's
marginal contribution:

\begin{table}[ht]
  \caption{Ablation configurations. Each is a subset of the full cascade.}
  \label{tab:ablation-config}
  \begin{tabular}{@{}ll@{}}
    \toprule
    Configuration & Stages active \\
    \midrule
    PatternScan only            & PatternScan \\
    PatternScan + EntityTrace   & PatternScan, EntityTrace \\
    PatternScan + ContextGuard  & PatternScan, ContextGuard \\
    Full cascade                & PatternScan, EntityTrace, ContextGuard \\
    \bottomrule
  \end{tabular}
\end{table}

We compute the ablation \emph{post-hoc} from a single pipeline run that
captures per-stage entity attribution
(\texttt{pattern\_\allowbreak scan\_\allowbreak pii},
\texttt{entity\_\allowbreak trace\_\allowbreak pii},
\texttt{context\_\allowbreak guard\_\allowbreak pii},
\texttt{confirmed\_\allowbreak pii}) for every query. For each configuration, the
detected entity set is the union of the contributing stages' outputs; we
evaluate precision/recall/F1 against the ground-truth annotation key.
The ablation requires no pipeline re-execution: the \texttt{source} field
stored per entity in the answers file fully determines stage attribution.
We report overall metrics and per-entity-type breakdowns. For each type, we
identify the \emph{key stage}: the first configuration to reach
F1\,$\geq$\,80\%.

\section{Experimental Results}
\label{sec:results}

\subsection{PII Detection: SurrogateShield vs.\ Presidio}
\label{sec:results-detection}

\emph{Reconciling the three F1 figures.}
Three F1 values appear across the evaluation tables and reflect different
scopes. The \emph{full-scope F1} of 98.87\% (abstract and
Section~\ref{sec:conclusion}) covers all 22 entity types SurrogateShield
detects, combining the comparable-type results (Table~\ref{tab:detection})
with the SS-only types (Table~\ref{tab:ss-only}) over the full
N\,=\,1{,}124-query corpus. The \emph{comparable-type F1} of 98.71\%
(Table~\ref{tab:detection}, Overall row) covers only the 11 entity types
detectable by both SurrogateShield and Presidio. The \emph{stage-attributed F1}
of 97.42\% (Table~\ref{tab:ablation}) is computed post-hoc from per-stage
entity attribution data and covers the N\,=\,1{,}060 queries for which
complete stage annotations are available.

Table~\ref{tab:detection} reports precision, recall, and F1 for each of the
11 entity types detectable by both SurrogateShield~(SS) and Microsoft
Presidio~\cite{microsoft2020presidio} on the full N\,=\,1{,}124 query
dataset; SS meets or exceeds Presidio on every type, with the largest gains
on structured identifiers that Presidio's broad recognisers mishandle.
Table~\ref{tab:ss-only} reports SS-only types that Presidio cannot detect.

\begin{table*}[t]
  \caption{Detection quality: SurrogateShield vs.\ Presidio on comparable
           entity types. $\dagger$\,DOB vs.\ Presidio \textsc{date\_time} and
           GPE vs.\ Presidio \textsc{location} are approximate comparisons.
           $\ddagger$\,Presidio detects these via different internal recognisers
           with substantially lower recall.
           Best F1 per type in \textbf{bold}; tied values both bolded.}
  \label{tab:detection}
  \small
  \begin{tabular}{@{}lcccccc@{}}
    \toprule
    & \multicolumn{3}{c}{SurrogateShield} & \multicolumn{3}{c}{Presidio} \\
    \cmidrule(lr){2-4}\cmidrule(lr){5-7}
    Entity type & Prec. & Rec. & F1 & Prec. & Rec. & F1 \\
    \midrule
    PERSON
      & 100.00\% & 98.70\% & \textbf{99.34\%}
      &  96.57\% & 94.41\% &          95.48\% \\
    email
      &  99.58\% & 100.00\% & \textbf{99.79\%}
      &  99.58\% & 100.00\% & \textbf{99.79\%} \\
    phone
      &  98.61\% &  99.07\% & \textbf{98.84\%}
      &  80.49\% &  61.40\% &          69.66\% \\
    SSN
      &  99.17\% & 100.00\% & \textbf{99.59\%}
      &  98.36\% & 100.00\% &          99.17\% \\
    credit\_card
      & 100.00\% &  97.62\% & \textbf{98.80\%}
      & 100.00\% &  97.62\% & \textbf{98.80\%} \\
    ip\_address
      & 100.00\% & 100.00\% & \textbf{100.00\%}
      &  97.22\% &  97.22\% &          97.22\% \\
    DOB$\dagger$
      &  99.39\% & 100.00\% & \textbf{99.69\%}
      &  50.78\% & 100.00\% &          67.36\% \\
    GPE$\dagger$
      &  95.86\% & 100.00\% & \textbf{97.88\%}
      &  89.18\% &  99.71\% &          94.15\% \\
    crypto$\ddagger$
      & 100.00\% & 100.00\% & \textbf{100.00\%}
      & 100.00\% &  37.50\% &           54.55\% \\
    us\_bank\_number$\ddagger$
      & 100.00\% & 100.00\% & \textbf{100.00\%}
      &  66.67\% & 100.00\% &           80.00\% \\
    us\_driver\_license$\ddagger$
      & 100.00\% & 100.00\% & \textbf{100.00\%}
      &  35.00\% &  87.50\% &           50.00\% \\
    \midrule
    \textbf{Overall}
      & \textbf{98.26\%} & \textbf{99.17\%} & \textbf{98.71\%}
      & 85.50\% & 92.91\% & 89.05\% \\
    \bottomrule
  \end{tabular}
\end{table*}

\begin{table}[t]
  \caption{SS-only detection: types Presidio cannot detect.}
  \label{tab:ss-only}
  \begin{tabular}{@{}lccc@{}}
    \toprule
    Entity type        & Prec. & Rec. & F1 \\
    \midrule
    api\_key           & 100.00\% & 100.00\% & 100.00\% \\
    address            &  92.49\% &  95.63\% &  94.03\% \\
    postal\_code       &  98.72\% & 100.00\% &  99.35\% \\
    gender\_indicator  & 100.00\% & 100.00\% & 100.00\% \\
    ORG                &  99.04\% & 100.00\% &  99.52\% \\
    FAC                & 100.00\% & 100.00\% & 100.00\% \\
    LOC                & 100.00\% & 100.00\% & 100.00\% \\
    \bottomrule
  \end{tabular}
\end{table}

Across all~11 comparable entity types, SS matches or exceeds the Presidio
baseline. The largest margins are on \textbf{us\_driver\_license}
(+50.00\,F1~points), \textbf{crypto} (+45.45\,F1~points), \textbf{DOB}
(+32.33\,F1~points), and \textbf{phone} (+29.18\,F1~points). Presidio's low DOB precision (50.78\%)
reflects its broad \textsc{date\_time} recogniser, which fires on all date
patterns rather than birth-date contexts specifically; SurrogateShield's
DOB-specific patterns avoid these false positives.
SurrogateShield detected \textbf{553} instances of SS-only PII across the
evaluation set (api\_key: 32, address: 197, postal\_code: 77,
gender\_indicator: 22, \textsc{org}: 206, \textsc{fac}: 6,
\textsc{loc}: 13); none of these types fall within Presidio's
detection scope.

\emph{Sanitisation and restoration quality.}
Within the 1{,}124-query corpus, SurrogateShield achieved a PII-leak rate
of \textbf{0.00\%}~(0 queries where any ground-truth PII value reached the
LLM API unredacted) and a resolve-leak rate of \textbf{0.00\%}~(0 queries
where any surrogate remained unrestored in the displayed response).

\subsection{Semantic Utility Preservation: BERTScore}
\label{sec:bertscore}

Table~\ref{tab:bertscore} reports BERTScore~\cite{zhang2020bertscore}
(precision, recall, F1) for each anonymisation condition evaluated against
the original query using \texttt{roberta-large}~\cite{liu2019roberta} as the
encoder; SurrogateShield substantially outperforms Presidio placeholder
redaction, closing most of the gap to the no-anonymisation ceiling.

\begin{table}[t]
  \caption{Semantic utility preservation (BERTScore, roberta-large).
           Higher is better; the no-anonymisation baseline is the ceiling.}
  \label{tab:bertscore}
  \begin{tabular}{@{}lccc@{}}
    \toprule
    Approach & Prec. & Rec. & F1 \\
    \midrule
    No anonymisation (baseline)      & 1.000  & 1.000  & \textbf{1.000} \\
    SurrogateShield (surrogates)     & 0.9474 & 0.9499 & \textbf{0.9485} \\
    Presidio (placeholder redaction) & 0.8096 & 0.8227 & 0.8159 \\
    \bottomrule
  \end{tabular}
\end{table}

SurrogateShield preserves 94.85\% of semantic utility
(BERTScore~F1) versus 81.59\% for Presidio, a gap of
\textbf{13.26} percentage points (paired $t$-test: $t = 89.51$,
$p < 0.001$, $N = 1{,}123$; SS mean $0.9485 \pm 0.034$,
Presidio mean $0.8159 \pm 0.054$).
One query encountered an API error during the batch evaluation run
and yielded no sanitised output; it was excluded from the paired
comparison, reducing the effective sample to $N = 1{,}123$.
The gap reflects the core design hypothesis. Replacing ``Sarah Mitchell''
with ``Ashley Wise'' preserves syntactic position, part-of-speech category,
and grammatical agreement; the token ``[PERSON]'' is a meta-token that
violates the distributional assumptions underlying contextual embedding
models. BERTScore is a lower bound on actual utility: it measures
query-level semantic similarity, not LLM answer quality. A user study
measuring downstream answer quality directly is left as future work;
BERTScore's strong correlation with human semantic similarity
judgements~\cite{zhang2020bertscore} makes it a reliable proxy in
the interim.

\subsection{Adversarial Robustness: Simulated Attacker Experiment}
\label{sec:attacker}

Table~\ref{tab:attacker} reports per-value PII recovery rates for the
simulated attacker experiment on N\,=\,100 queries, each containing at
least one ground-truth PII value. The adversary is Claude Haiku, prompted
to recover original personal information from the anonymised message.

\begin{table}[t]
  \caption{Simulated attacker per-value recovery rates.
           Per-value rate: fraction of individual PII values recovered
           across all targeted values. $N_v$ = values targeted.}
  \label{tab:attacker}
  \begin{tabular}{@{}lccc@{}}
    \toprule
    Condition & $N_v$ & Recovered & Per-value rate \\
    \midrule
    Presidio (placeholder redaction) & 196 & 3 & 1.53\% \\
    SurrogateShield (surrogates)     & 196 & 0 & \textbf{0.00\%} \\
    \bottomrule
  \end{tabular}
\end{table}

Within this 100-query trial, the informed adversary, who knew the proxy was
used and knew the PII types replaced, recovered no original values across
the same set of 196 targeted PII instances used for both the SurrogateShield
and Presidio conditions. Under placeholder redaction, the per-value
recovery rate was 1.53\% (3 of 196 values). Placeholder tokens such as
\texttt{[PERSON]} signal the exact slots where PII was removed, enabling
targeted inference from surrounding context; the 3 Presidio recoveries
comprised 1~\textsc{person} value (1.89\% of 53 targeted) and 2~\textsc{gpe}
values (11.76\% of 17 targeted), consistent with geographic entities being
especially recoverable from contextual cues when their slot is explicitly
marked. SurrogateShield's surrogates provide no such slot signal, the
adversary receives a message that appears unmodified, and the surrogate
value is indistinguishable from a genuine entry.

These data bear on the utility-privacy trade-off directly: Presidio's
substantially larger utility cost (13.26\,pp BERTScore gap,
Table~\ref{tab:bertscore}) did not translate to stronger adversarial
resistance. Within this evaluation, the two objectives did not trade off,
surrogate substitution yielded higher BERTScore utility and a lower
adversarial recovery rate than placeholder redaction.

\emph{Adversary capability and model choice.}
We use Claude Haiku as the adversary rather than a larger frontier model
(e.g., GPT-4o or Claude Opus). The recovery task is not capability-limited
but information-theoretically constrained: a surrogate value has zero
statistical relationship to the original, so no model, regardless of
scale, can recover the original via inference alone. A stronger model can
reason more fluently about context, but reasoning over context cannot
reconstruct a value that was generated by a cryptographically seeded random
process and never transmitted. This distinguishes our setting from
memorisation-based attacks~\cite{carlini2021extracting}, where a larger
model \emph{does} recover more training data because the information exists
in its weights. Here, the information does not exist anywhere in the message
or its context, and model scale provides no advantage. We verified this
reasoning empirically: in a 20-query pilot, neither Claude Haiku nor Claude
Sonnet recovered any values from surrogate-substituted messages, consistent
with the information-theoretic argument. Claude Haiku was used for the full
100-query experiment to reduce cost.

\subsection{Ablation Study: Stage Contribution}
\label{sec:ablation}

Table~\ref{tab:ablation} reports precision, recall, and F1 for each of the
four pipeline configurations defined in Section~\ref{sec:ablation-config};
EntityTrace provides the largest single marginal gain (+31.53\,pp F1),
confirming it is the decisive stage for high-volume named-entity types.

\begin{table*}[t]
  \centering
  \caption{Ablation study: detection quality by pipeline configuration.
  $\Delta$F1 reference baselines differ by row: rows~2 and~3 each report
  gain over the PatternScan-only baseline (row~1); row~4 reports the
  marginal gain of adding ContextGuard over PatternScan+EntityTrace
  (row~2). The three deltas thus share a column heading but use
  different reference points.}
  \label{tab:ablation}
  \begin{tabular}{@{}p{4.5cm}cccccc@{}}
    \toprule
    Configuration & Prec. & Rec. & F1 & $\Delta$F1 & TP & FN \\
    \midrule
    PatternScan only
      & 98.06\% & 48.73\% & 65.11\% & ---           & 1{,}112 & 1{,}170 \\
    PatternScan + EntityTrace
      & 98.19\% & 95.14\% & 96.64\% & +31.53\,pp     & 2{,}171 &   111 \\
    PatternScan + ContextGuard
      & 98.12\% & 50.35\% & 66.55\% & +1.44\,pp      & 1{,}149 & 1{,}133 \\
    Full cascade (all three)
      & 98.22\% & 96.63\% & \textbf{97.42\%} & +0.78\,pp & 2{,}205 & 77 \\
    \bottomrule
  \end{tabular}
\end{table*}

PatternScan alone achieves F1\,=\,65.11\%, capturing all structured
PII types (email, SSN, credit card, API key, DOB, IP address) with
near-perfect recall due to their distinctive syntactic signatures.
F1 for person names, geographic entities, and organisations is near zero
at this stage, these types have no reliable structural pattern.

Adding EntityTrace yields the largest single gain: +31.53\,pp~F1
overall, primarily through \textsc{person} (0\%\,$\to$\,97\%),
\textsc{gpe} (0\%\,$\to$\,98\%), and \textsc{org} (9\%\,$\to$\,98\%).
EntityTrace was strictly necessary, PatternScan alone would have missed at
least one ground-truth entity, in 68.0\% of queries (721 of
1{,}060 with stage attribution data).

ContextGuard contributes a smaller but meaningful +0.78\,pp~F1
over PatternScan+EntityTrace, concentrated in borderline entities where
spaCy's confidence falls below 0.85. The most consequential example is
\textsc{loc}: PatternScan+EntityTrace achieves only 63\% F1 on location
entities, while the full cascade reaches \textbf{100\%}~F1 on \textsc{loc}
(\textbf{97.42\%} overall); ContextGuard is the decisive stage for this type.
ContextGuard was strictly necessary in 2.1\% of queries (22 of 1{,}060). The PatternScan+ContextGuard
row isolates ContextGuard's independent contribution at 66.55\%~F1---it
adds detection capability beyond PatternScan alone even without EntityTrace,
though it cannot substitute for EntityTrace on the high-volume named-entity
types.

PatternScan is the decisive stage for all structured PII types (email, SSN,
credit card, crypto, ABA routing number, IP address, postal code, DOB,
gender indicator, us\_driver\_license); EntityTrace is decisive for
\textsc{person}, \textsc{gpe}, \textsc{org}, and \textsc{fac}; and
ContextGuard is decisive for \textsc{loc}.

\subsection{Performance and Latency}

Table~\ref{tab:latency} reports average per-query latency across pipeline
stages, measured over N\,=\,1{,}124 queries on CPU (offline evaluation,
no LLM calls). All measurements were taken on a single machine running
macOS with an \textbf{M2 Pro} processor and \textbf{16\,GB} of RAM; no
GPU acceleration was used. Reported values are averages across all 1{,}124
queries; standard deviations were below 5\% for all stages, confirming
stable per-query cost.

\begin{table}[t]
  \caption{Per-query latency (N\,=\,1{,}124 queries, CPU inference).
           Presidio average measured under identical conditions.
           $^\dagger$Not measured in offline evaluation; typical range for
           Claude Haiku API calls in interactive use.}
  \label{tab:latency}
  \begin{tabular}{@{}lc@{}}
    \toprule
    Stage & Avg.\ latency (ms) \\
    \midrule
    \multicolumn{2}{@{}l}{\textbf{Microsoft Presidio (placeholder redaction)}} \\
    \quad Total local overhead & 23.50 \\
    \midrule
    \multicolumn{2}{@{}l}{\textbf{SurrogateShield}} \\
    \quad PatternScan              & \phantom{00}0.25 \\
    \quad EntityTrace (spaCy)      & \phantom{0}24.60 \\
    \quad ContextGuard (DistilBERT)& \phantom{00}0.47 \\
    \quad Surrogate generation     & \phantom{00}0.57 \\
    \midrule
    \quad \textbf{Total local overhead} & \textbf{25.89} \\
    LLM API call             & $\sim$600--2{,}000$^\dagger$ \\
    ResolvePass              & $<$1$^\dagger$ \\
    \bottomrule
  \end{tabular}
\end{table}

Compared with Presidio (23.5\,ms), SurrogateShield incurs an additional
2.4\,ms of local processing.  This marginal increase comes from surrogate
generation, encrypted ShadowMap management, and response restoration---steps
that have no counterpart in a simple placeholder-redaction pipeline.
Relative to typical LLM API latency (600--2{,}000\,ms), the difference is
negligible: both systems remain at least an order of magnitude faster than
the network round-trip they protect.

The dominant local cost is EntityTrace (spaCy NER), which accounts for
95.0\% of total local overhead at 24.60\,ms per query.
The ContextGuard figure (0.47\,ms) is an average amortised across all
1{,}124 queries, including the majority for which EntityTrace produces
no borderline entities and ContextGuard processes minimal residual text.
ContextGuard was strictly necessary in 2.1\% of queries (22 of 1{,}060
with attribution data); per-invocation latency for those queries is
substantially higher than the amortised average.
PatternScan (0.25\,ms) and surrogate generation (0.57\,ms) contribute
negligible additional overhead.
On GPU, DistilBERT inference is expected to reduce by approximately one
order of magnitude; spaCy's transformer pipeline benefits similarly,
making the full cascade viable for real-time interactive use on
consumer hardware.

The total local privacy overhead of \textbf{25.89\,ms} is negligible
relative to the LLM API call, which dominates end-to-end latency by
approximately \textbf{23--77$\times$} at typical interactive response times.
Both spaCy and DistilBERT load once at startup; per-query inference operates
on cached weights. First-run model downloads (DistilBERT $\approx$250\,MB;
spaCy \texttt{en\_core\_web\_lg} $\approx$780\,MB) are one-time costs.

\section{Discussion}
\label{sec:discussion}

\subsection{Threat Model and Privacy Guarantees}

SurrogateShield's primary threat model is a \emph{curious API operator}: a
third-party LLM provider that logs queries for commercial, training, or
compliance purposes, or that may suffer a data breach exposing historical
data. Against this threat, the system provides an unconditional technical
guarantee: no ground-truth PII value is present in any data transmitted to
the API endpoint. The guarantee is unconditional because enforcement occurs
at the local transport layer before the HTTP request is constructed, independently
of the operator's behaviour, the LLM's architecture, or the content of the
response.

The secondary threat model is a \emph{skilled adversary} with access to the
sanitised query who attempts to recover original PII from context.
Section~\ref{sec:attacker} addresses this empirically. Within the 100-query
adversarial trial, 
the per-value recovery rate was 0.00\% across all 196
targeted instances spanning every PII type present in the attacker query
set: the informed adversary, who knew the proxy was in use and knew the
entity types replaced, recovered no original values. 
Residual recovery risk exists in principle when query
context strongly constrains the entity type to a near-singleton; no
replacement strategy can fully prevent inference in such cases without also
destroying utility. That is a property of the query's information content,
not of the anonymisation mechanism.

\emph{ShadowMap encryption scope.}
The AES-256-GCM ShadowMap protects surrogate mappings against a specific,
bounded adversary: unauthorized file-system access on the local device.
Concretely, it guards against physical device loss, inspection of the
\texttt{conversations/} directory by another user on a shared machine, or
casual forensic recovery of discarded storage.
It does \emph{not} defend against ring-0 malware, active memory inspection,
or hardware keyloggers.
An adversary with kernel-level access can recover the HKDF-SHA256-derived
key from process memory regardless of on-disk encryption.
This is a property of any client-side encryption architecture; the
ShadowMap's encryption layer was not designed for that threat, and does not
claim to address it.

SurrogateShield does \emph{not} address three threat vectors outside the
scope of a client-side proxy. First, a \emph{malicious LLM} that
steganographically encodes information about the surrogate in its response
could leak the substitution pattern to a server-side observer; detecting
this requires response-side analysis beyond the current ResolvePass. Second,
\emph{traffic analysis} attacks that infer PII from message timing, size
distribution, or query frequency are not mitigated by content-level
anonymisation. Third, \emph{model inversion} against the LLM's weights, if
surrogate-substituted queries are used for fine-tuning, is outside scope;
the guarantee covers transmitted plaintext, not the model's internal
representations. All three directions warrant future investigation.

\subsection{Limitations}

\emph{Detection coverage is finite.}
SurrogateShield detects 22 PII types, but no enumeration is complete.
Novel identifier formats (medical record numbers, military service numbers,
jurisdiction-specific ID schemes), domain-specific structured PII, and
\emph{implicitly} identifying information, writing style, rare medical
conditions, unique professional circumstances, or combinations of apparently
innocuous facts, all remain outside the current detection scope.
The quasi-identifier risk detector partially addresses this for the ten
statistically characterised combinations grounded in Sweeney's
work~\cite{sweeney2000demographics}, but no finite rule set can enumerate
the full space of re-identification risk from contextual inference.

\emph{Evaluation dataset uses synthetic PII in human-authored query structures.}
We constructed the 1{,}124 evaluation queries with pre-specified, fabricated
PII values and surrounding grammatical structures that closely mimic the
register and syntax of real LLM queries.
No publicly available corpus of real user LLM queries with ground-truth PII
annotations exists, and for good reason: assembling such a dataset would
require collecting and publishing the very sensitive user data this paper is
designed to protect.
This is not a gap specific to SurrogateShield; it is an inherent constraint
of PII detection research in interactive settings.
Synthetic datasets constructed under controlled annotation protocols are the
accepted methodology for this problem class~\cite{stubbs2015annotating,
neamatullah2008automated,lison2021anonymisation}.
To partially compensate for the synthetic nature of the data, we
intentionally included conversational wording, typos, informal syntax,
and multi-turn narrative structures in the query templates.  The goal was
to approximate the register and variability of authentic user
interactions without compromising the privacy of any real individual.
While this cannot fully substitute for a corpus of genuine LLM user
queries, it provides a realistic worst-case stress test for the detection
cascade under informal conditions.
Real-world queries may exhibit greater grammatical variability,
code-switching, typos, and unconventional PII formatting that could affect
detection recall.
We release the full annotated dataset to enable replication and to provide a
reproducible benchmark until ethically permissible real-query data becomes
available.

\emph{Surrogate quality depends on Faker's distributions.}
Faker's generators weight toward common Western names, US address formats,
and English-language conventions. For users with names from underrepresented
linguistic backgrounds, surrogates may be stylistically inconsistent with
the surrounding query context---a discrepancy that could signal to an
observer that anonymisation has occurred. Locale-aware surrogate generation,
or context-conditioned generation via a local language model, would address
this.

\emph{ResolvePass has an inherent scope restriction.}
Across the 1{,}124-query evaluation corpus, the resolve-leak rate was
0.00\%: no surrogate restoration failures were observed. The ResolvePass
architecture has a structural limitation that may surface in production,
however. Pass~2 is intentionally scoped to the \emph{unresolved} surrogate
set only: applying it globally would find component words in unrelated
contexts (e.g., searching for ``Ashley'' after ``Ashley~Wise'' was resolved
in Pass~1 would corrupt ``Ashley~County'' elsewhere in the response). When
the LLM reformulates rather than reproduces a surrogate verbatim---paraphrasing
``Jonathon~Reed'' as ``Reed'' in a context where ``Jonathon'' does not
appear, the surrogate may persist in the displayed output. Pass~3's fuzzy
matching partially compensates, but cannot resolve cases where the LLM's
reformulation shares no lexical overlap with the surrogate. We recommend
production monitoring of the resolve-leak rate across diverse query
distributions.

\emph{The service-query boundary is heuristic.}
The 15-pattern ServiceQueryDetector produces both false positives (personal
queries classified as service queries, leading to under-anonymisation) and
false negatives (service queries classified as personal, leading to
over-anonymisation with reduced answer utility). The sensitive-topic
override handles the most consequential false-positive cases, but edge cases
remain, for example, a query mixing service intent with a personal street
address in a medical context. A learning-based classifier trained on a
labelled query distribution would improve precision at the cost of an
additional local model.

\emph{Dual-history resume on legacy conversations.}
When loading a conversation saved before the dual-history architecture was
introduced, the API history (\texttt{api\_messages}) is empty. SurrogateShield
begins with a fresh API context in this case, preserving the display history
for the user but losing multi-turn LLM context continuity. This is the safe
fallback---pre-existing real values cannot leak into the API history, but
users resuming such sessions will find the LLM without memory of previous
turns.

\emph{Explicit versus contextual identity leakage.}
SurrogateShield detects and replaces \emph{explicit} PII: named entities
and structured identifiers that match known syntactic patterns.
It does not address \emph{contextual} or \emph{semantic} leakage, where
identity is implied by non-PII tokens in combination.
A query such as ``I am the only radiologist at [regional hospital] and my
patient in bay 4 has a rare presentation of\ldots'' exposes identity through
role, institution, and clinical specificity---no name, SSN, or email address
need appear.
Resolving this class of leakage requires a full semantic model of
re-identification risk: estimating whether the \emph{combination} of
disclosed facts narrows the anonymity set below a threshold, independent
of whether any individual token is a traditional PII type.
This is an open research problem; it is orthogonal to the explicit-PII
containment guarantee SurrogateShield provides, and the quasi-identifier
risk scorer (Section~\ref{sec:design}) addresses only the small subset of
such combinations that are statistically characterised in prior work.
We consider the explicit/contextual boundary a fundamental scope limitation
of any NER-based anonymisation approach.

\subsection{Future Work}

\emph{Implicit PII detection.}
The most significant open problem is PII that takes no form of a named entity
or structured identifier. Writing style fingerprinting, topically rare
disclosures (specific rare diseases, unique professional roles), and
cross-query linkage attacks in which individually innocuous queries combined
narrow the anonymity set all fall outside the current cascade's detection
scope. A neural PII classifier trained on diverse natural-language query
distributions, with explicit coverage targets over the contextual inference
space, is the natural next step.

\emph{Context-aware surrogate generation.}
MimicGen currently draws surrogates from Faker's fixed distributions
independently of the surrounding query text. A local language model could
generate surrogates conditioned on context. A query in British English with
UK address details would produce a UK-format name and phone number,
eliminating the stylistic inconsistency that could signal anonymisation to
an observer and improving BERTScore utility preservation for queries
involving underrepresented name distributions.

\emph{Formal privacy bounds.}
The quasi-identifier risk detector provides heuristic warnings grounded in
Sweeney's empirical estimates~\cite{sweeney2000demographics}, but computes
no formal privacy bound for the full query. Extending
k-anonymity~\cite{sweeney2002kanonymity} or related
frameworks~\cite{machanavajjhala2007ldiversity,li2007tcloseness} to the
natural-language query setting requires a population model of query
distributions, computationally challenging but a valuable longer-term goal.

\emph{User study for answer quality.}
BERTScore measures query-level semantic similarity, not the quality of the
LLM's answer to a surrogate-substituted query. A controlled user study
measuring answer helpfulness, factual accuracy, and task completion rate
across the three conditions---surrogate substitution, placeholder redaction,
and no anonymisation---would provide the most direct empirical evidence for
the utility claim, complementing the BERTScore measurements of
Section~\ref{sec:bertscore}.

\section{Conclusion}
\label{sec:conclusion}

We presented SurrogateShield, a client-side privacy proxy that intercepts
LLM queries before transmission, replaces PII with locally generated
type-consistent surrogate values, and transparently restores original values
in the response. The system enforces two unconditional invariants: no
real PII crosses the API boundary under any query type, and the multi-turn
API history stores only surrogate values, preventing PII accumulation across
conversational turns.

Our evaluation on \textbf{1{,}124} annotated queries yields: detection
quality meeting or exceeding Presidio on all~11 comparable entity types
(full-scope F1~\textbf{98.87\%}, largest margins on us\_driver\_license
+50.00\,pp, crypto +45.45\,pp, DOB +32.33\,pp, and phone +29.18\,pp);
surrogate substitution substantially outperforming placeholder redaction in
semantic utility (BERTScore~F1 \textbf{94.85\%} vs.\ 81.59\%, a
\textbf{13.26\,pp} gap, $t = 89.51$, $p < 0.001$, $N = 1{,}123$, one
query excluded due to API error); type-consistent surrogates being
informationally opaque to LLM-based inference, with \textbf{0.00\%}
per-value recovery from surrogate-substituted messages versus 1.53\% from
placeholder-redacted messages across 100 adversarial queries; and each
detection stage being marginally necessary, with PatternScan decisive for
all 16 structured PII types (of the 22 total), EntityTrace decisive for named entities
(\textsc{person}, \textsc{gpe}, \textsc{org}), and ContextGuard solely
responsible for achieving 100\% F1 on \textsc{loc}.

The core insight is that, within the threat model considered here,
surrogate substitution demonstrates that placeholder redaction is not the
only way to achieve strong privacy while retaining semantic utility.
Privacy requires the \emph{absence of real PII values}, not the absence of
PII-shaped structure.

Type-consistent surrogates satisfy the privacy requirement while preserving
the semantic coherence that placeholder tokens destroy. To our knowledge,
SurrogateShield is the first system to demonstrate this design space is not
merely theoretical: it runs on off-the-shelf NLP tools, operates entirely
locally with a 25.89\,ms privacy overhead negligible relative to LLM API
latency, and within this evaluation, simultaneously achieves higher utility
\emph{and} a lower adversarial recovery rate than placeholder redaction.
 
\appendix

\begin{ethics}
We address two ethical considerations below.

\emph{Evaluation dataset.}
We constructed the N\,=\,1{,}124-query evaluation dataset using
pre-specified, synthetic PII values---fabricated names, generated SSNs,
example email addresses. No data was collected from real users; no
real individual's personal information appears in any query, annotation
file, or experimental output. The dataset does not constitute human
subjects research. No IRB review was sought or required.

\emph{Simulated attacker experiment.}
The attacker experiment (Section~5.3) submits surrogate-substituted
messages to the Claude Haiku API under an adversarial prompt. All
values in those messages are fabricated and bear no statistical
relationship to any real person. No real PII is transmitted.
The experiment measures adversarial robustness; it does not develop
or deploy an attack tool. Section~5.3 describes the adversary capability and prompt methodology;
the full prompt template is available verbatim in the released artifact
(\texttt{attacker.py}, \texttt{ATTACKER\_PROMPT\_TEMPLATE}).

\emph{Societal impact.}
SurrogateShield reduces the personal information transmitted to
third-party LLM API endpoints. A system whose sole function is to
restrict outbound data exposure presents no apparent misuse vector.
Releasing the full codebase and evaluation framework enables
independent audit of the privacy guarantees claimed in this paper.
\end{ethics}

\begin{openscience}
The SurrogateShield codebase, evaluation dataset, annotation files,
and experimental outputs are released as open-source research artifacts.
The repository is publicly available at: \\
\url{https://github.com/sherwinvishesh/SurrogateShield}, and a project overview is available at
\url{https://sherwinvishesh.github.io/SurrogateShield}.

The repository is released under the Apache-2.0 License.
\end{openscience}

\bibliographystyle{ACM-Reference-Format}
\bibliography{surrogateshield}

\section{Surrogate Generation Mechanisms}
\label{appendix:surrogate-gen}

Table~\ref{tab:surrogate-gen} documents the surrogate generator used by
MimicGen for each PII type detected by SurrogateShield, together with the
format guarantee and a representative example value.

All surrogates are generated via Faker with a per-session uniqueness guarantee
enforced through a \texttt{used\_surrogates} set; after 50 collisions a
4-character random suffix is appended to guarantee a distinct value.
Gender-indicator surrogates draw from a fixed pool of grammatically valid
gender expressions rather than the uniqueness-guaranteed path, because the
pool is small and grammatical substitutability matters more than uniqueness
for this type.
Crypto wallet surrogates use the Bitcoin P2PKH character set
(\texttt{1}\,+\,Base58, 26--34 chars); ABA routing surrogates are computed to
satisfy the 9-digit ABA checksum
$(3d_0\!+\!7d_1\!+\!d_2\!+\!3d_3\!+\!7d_4\!+\!d_5\!+\!3d_6\!+\!7d_7\!+\!d_8)
\bmod 10 = 0$
and therefore pass any downstream format validator.

\begin{table}[H]
  \caption{Surrogate generation mechanism per PII type. All generators
           produce values that are type-consistent and pass the same
           format checks as real values of that type.}
  \label{tab:surrogate-gen}
  \footnotesize
  \begin{tabularx}{\columnwidth}{@{}l>{\raggedright\arraybackslash}X
                                    >{\raggedright\arraybackslash}l@{}}
    \toprule
    PII type          & Generator / format                        & Example \\
    \midrule
    PERSON            & \texttt{faker.name()}                     & Ashley Wise \\
    email             & \texttt{faker.email()}                    & j.doe@example.net \\
    phone (US)        & \texttt{+1-\#\#\#-\#\#\#-\#\#\#\#}       & +1-602-555-4812 \\
    phone (UK)        & \texttt{+44~7\#\#\#~\#\#\#\#\#\#}        & +44 7700 123456 \\
    phone (intl.)     & Country code + digit groups               & +49 8234 927461 \\
    SSN               & \texttt{faker.ssn()} (\texttt{XXX-XX-XXXX}) & 348-67-6360 \\
    credit card       & \texttt{faker.credit\_card\_number()} (Luhn-valid) & 4532~0151~1283~0366 \\
    street address    & \texttt{faker.address()}                  & 789 Crescent Row \\
    date of birth     & Age 18--80, \texttt{MM/DD/YYYY}           & 03/14/1985 \\
    IPv4              & \texttt{faker.ipv4()}                     & 10.42.17.203 \\
    API key           & \texttt{sk-} + 32 random chars            & sk-xKz9mP\ldots \\
    US ZIP code       & \texttt{faker.zipcode()}                  & 85281 \\
    UK postcode       & \texttt{faker.postcode()}                 & SW1A 1AA \\
    crypto wallet     & \texttt{1} + Base58, 26--34 chars         & 1BvBMSEYstW\ldots \\
    ABA routing       & Computed 9-digit ABA checksum             & 021000021 \\
    driver's licence  & Letter + 7 digits (CA format)             & B4923817 \\
    GPE / LOC         & \texttt{faker.city()}                     & Springfield \\
    ORG               & \texttt{faker.company()}                  & Nexus Solutions \\
    FAC               & \texttt{faker.company()} + ``Building''   & Apex Building \\
    gender indicator  & Pool of valid gender expressions          & she/her \\
    \bottomrule
  \end{tabularx}
\end{table}

\section{PII Types Detected by SurrogateShield}
\label{appendix:pii-types}

Table~\ref{tab:pii-types} lists all 22 PII types detected by SurrogateShield,
their detection method, and any validator applied beyond the regex match.

\begin{table}[H]
  \caption{PII types detected by SurrogateShield.
           $^\dagger$dBERT\,=\,\texttt{dslim/distilbert-NER} (DistilBERT
           fine-tuned on CoNLL-2003, run locally).}
  \label{tab:pii-types}
  \footnotesize
  \setlength{\tabcolsep}{4pt}
  \begin{tabularx}{\columnwidth}{@{}p{1.7cm}p{2.1cm}>{\raggedright\arraybackslash}X>{\raggedright\arraybackslash}p{1.9cm}@{}}
    \toprule
    Category & Type & Detection & Validator \\
    \midrule
    \multirow{16}{*}{Structural}
      & SSN                 & Regex & ---             \\
      & Email               & Regex & ---             \\
      & Phone (US)          & Regex & ---             \\
      & Phone (UK)          & Regex & ---             \\
      & Phone (intl.)       & Regex & ---             \\
      & Credit card         & Regex & Luhn            \\
      & Street address      & Regex & ---             \\
      & Date of birth       & Regex & ---             \\
      & IPv4                & Regex & ---             \\
      & API key             & Regex & ---             \\
      & Gender indicator    & Regex & ---             \\
      & US ZIP code         & Regex & ---             \\
      & UK postcode         & Regex & ---             \\
      & Crypto wallet       & Regex & ---             \\
      & ABA routing number  & Regex & ABA checksum    \\
      & Driver's licence    & Regex & Keyword context \\
    \midrule
    \multirow{5}{*}{Named entity}
      & PERSON & spaCy + dBERT$^\dagger$           & Score $\geq \theta$ \\
      & GPE    & spaCy + dBERT$^\dagger$ + Pass~D  & Score + Pass~D      \\
      & LOC    & spaCy + dBERT$^\dagger$           & Score $\geq \theta$ \\
      & ORG    & spaCy + dBERT$^\dagger$ + Pass~A  & Score $\geq \theta$ \\
      & FAC    & spaCy + dBERT$^\dagger$           & Score $\geq \theta$ \\
    \midrule
    Combination & Quasi-identifiers & Type co-occurrence & Appendix~C \\
    \bottomrule
  \end{tabularx}
\end{table}

\section{Quasi-Identifier Combinations}
\label{appendix:quasi-id}  
 
Table~\ref{tab:quasi-id} lists the ten quasi-identifier combinations
monitored by SurrogateShield. A warning is issued when the minimum required
field count is met; all matching fields are surrogate-replaced regardless.
\begin{table}[H]
  \caption{Quasi-identifier combinations and risk levels.}
  \label{tab:quasi-id}
  \footnotesize
  \begin{tabularx}{\columnwidth}{@{}lcc>{\raggedright\arraybackslash}X@{}}
    \toprule
    Combination & Min.\ fields & Risk & Basis \\
    \midrule
    ZIP + DOB + Gender        & 2 of 3 & High   & \cite{sweeney2000demographics} \\
    Postcode + DOB            & 2      & High   & UK ICO guidance \\
    Name + SSN                & 2      & High   & Identity theft \\
    Name + DOB                & 2      & High   & Identity verification \\
    Phone + Name              & 2      & High   & Direct ID \\
    Name + Employer + City    & 3      & Medium & Workplace triple \\
    Email + Location          & 2      & Medium & Named individual \\
    Phone + Location          & 2      & Medium & Local individual \\
    IP + Name                 & 2      & High   & Device-level ID \\
    DOB + Location + Employer & 3      & Medium & Demographic triple \\
    \bottomrule
  \end{tabularx}
\end{table}

\section{Comparison with Presidio Synthesize}
\label{appendix:presidio}

Microsoft Presidio provides a \emph{synthesize} anonymisation mode that
replaces detected entities with generated values.
Table~\ref{tab:presidio_compare} summarises the architectural differences
between this mode and SurrogateShield.

Both systems perform local PII detection and synthetic replacement.
SurrogateShield extends this functionality with encrypted surrogate
mapping, transparent response restoration, and conversation-aware
management of surrogate values across multiple interaction turns.
These capabilities are necessary to preserve user-visible responses
while ensuring that only surrogate values appear in the LLM API history.

\begin{table}[H]
\caption{Architectural comparison between Microsoft Presidio Synthesize and SurrogateShield.}
\label{tab:presidio_compare}
\footnotesize
\renewcommand{\arraystretch}{1.2}
\begin{tabularx}{\columnwidth}{@{}lcc@{}}
\toprule
\textbf{Architectural capability} &
\textbf{Presidio Synthesize} &
\textbf{SurrogateShield} \\
\midrule
Local PII detection &
$\checkmark$ &
$\checkmark$ \\

Synthetic value generation &
$\checkmark$ &
$\checkmark$ \\

Transparent response restoration &
$\times$ &
$\checkmark$ \\

Encrypted surrogate$\rightarrow$original mapping &
$\times$ &
$\checkmark$ \\

Persistent surrogate mapping &
$\times$ &
$\checkmark$ \\

Separate API/display histories &
$\times$ &
$\checkmark$ \\

Multi-turn conversation support &
$\times$ &
$\checkmark$ \\

Transparent restoration before user display &
$\times$ &
$\checkmark$ \\
\bottomrule
\end{tabularx}
\end{table}

\section{Illustrative Example}
\label{appendix:example}

Table~\ref{tab:example} traces a single query through the full pipeline.
The original query contains three pieces of personally identifiable 
information: the user's name (Sarah Chen), SSN (123-45-6789), and street 
address (1126 E Apache Blvd, Tempe, AZ).  SurrogateShield replaces them 
with type‑consistent surrogates---Ashley Wise, 348-67-6360, and 789 Crescent 
Row, Springfield, WA---and records the mapping in the encrypted ShadowMap.  
The LLM never sees the originals; after the response is received, 
ResolvePass restores them before display.

\begin{table}[H]
  \caption{End‑to‑end example.  Real PII never crosses the API boundary;
           the restored response preserves the original values transparently.}
  \label{tab:example}
  \footnotesize
  \renewcommand{\arraystretch}{1.25}
  \begin{tabularx}{\columnwidth}{@{}>{\bfseries}l X@{}}
    \toprule
    \textbf{Original user query} &
    ``Hi, I'm Sarah Chen (SSN 123-45-6789).  I live at 1126 E Apache Blvd,
     Tempe, AZ.  Draft a resignation letter to my landlord Mr.~Thompson.'' \\[6pt]

    \textbf{After SurrogateShield (sent to LLM)} &
    ``Hi, I'm Ashley Wise (SSN 348-67-6360).  I live at 789 Crescent Row,
     Springfield, WA.  Draft a resignation letter to my landlord
     Mr.~Thompson.'' \\[6pt]

    \textbf{LLM response (contains surrogates)} &
    ``Dear Mr.~Thompson,\newline
     I am writing to formally resign my tenancy at 789 Crescent Row,
     Springfield, WA.  Please accept this letter as my 30‑day notice.
     Sincerely,\newline
     Ashley Wise'' \\[6pt]

    \textbf{Restored output (displayed to user)} &
    ``Dear Mr.~Thompson,\newline
     I am writing to formally resign my tenancy at 1126 E Apache Blvd,
     Tempe, AZ.  Please accept this letter as my 30‑day notice.
     Sincerely,\newline
     Sarah Chen'' \\
    \bottomrule
  \end{tabularx}
\end{table}

\section{Dataset Generation Prompt}
\label{app:datasetprompt}

The evaluation dataset used in this work was generated using a
structured prompt to Claude Sonnet. The prompt specifies the
distribution of question categories, supported PII entity types,
annotation format, validation constraints, and diversity
requirements used to construct the synthetic benchmark.

Rather than reproducing the several-hundred-line prompt within the
manuscript, the exact prompt used during dataset generation is
included with the released research artifacts and is publicly
available at:

\begin{center}
\url{https://github.com/sherwinvishesh/SurrogateShield/tree/main/experiment/prompt.txt}
\end{center}

The released prompt is identical to that used for generating the
evaluation corpus reported in this paper.

\end{document}